\documentclass{elsart}
\usepackage{amssymb,amsfonts,graphicx,bm}
\begin{document}
\begin{frontmatter}
\title{Transfer Entropy Analysis of the Stock Market}
\author{Seung Ki Baek},
\author{Woo-Sung Jung\corauthref{cor1}},
\ead{wsjung@kaist.ac.kr} \corauth[cor1]{Corresponding author. Fax:
+82-42-869-2510.}
\author{Okyu Kwon},
\author{Hie-Tae Moon}
\address{Department of Physics, Korea Advanced Institute of Science and Technology, Daejeon 305-701, Republic of Korea}

\begin{abstract}
In terms of transfer entropy, we investigate the strength and the
direction of information transfer in the US stock market. Through
the directionality of the information transfer, the more influential
company between the correlated ones can be found and also the market
leading companies are selected. Our entropy analysis shows that the
companies related with energy industries such as oil, gas, and
electricity influence the whole market.
\end{abstract}

\end{frontmatter}

\section{Introduction}
Recently, economy has become an active research area for physicists.
They have investigated stock markets using statistical tools, such
as the correlation function, multifractal, spin-glass models, and
complex networks
\cite{Arthur,Mantegna2000,Bouchaud2000,Mandelbrot2001,Kullmann2000,Giada2002}.
As a consequence, it is now found evident that the interaction
therein is highly nonlinear, unstable, and long-ranged.

All those companies in the stock market are interconnected and
correlated, and their interactions are regarded as the important
internal force of the market. The correlation function is widely
used to study the internal inference of the market
\cite{Mantegna1999,Onnela2003,Kullmann2002,Plerou1999,Jung2005}.
However, the correlation function has at least two limitations:
First, it measures only linear relations, although a linear model is
not a faithful representation of the real interactions in general.
Second, all it says is only that two series move together, and not
that which affects which: in other words, it lacks directional
information. Therefore participants located in hubs are always left
open to ambiguity: they can be either the most influential
ones or the weakest ones subject to the market trend all along. It
should be noted that introducing time-delay can be a good remedy for
these limitations. Some authors use such concepts as time-delayed
correlation and time-delayed mutual information, and these
quantities construct asymmetric matrices by preserving
directionality \cite{Kullmann2002,Schreiber1990}. In case that the
length of delay can be appropriately determined, one can also
measure the `velocity' whereby the influence spreads. In this paper,
however, we rely on a newly-devised variant of information to check
its applicability.

Information is an important keyword in analyzing the market
or in estimating the stock price of a given company. It is
quantified in rigorous mathematical terms \cite{Shannon}, and the
mutual information, for example, appears as meaningful choice
replacing a simple linear correlation even though it still does not
specify the direction. The directionality, however, is required to
discriminate the more influential one between correlated
participants, and can be detected by the transfer entropy (TE)
\cite{Schrb2000}.

This concept of TE has been already applied to the
analysis of financial time series by Marschinski and Kantz
\cite{Mars2002}. They calculated the information flow between the
Dow Jones and DAX stock indexes and obtained conclusions consistent
with empirical observations. While they examined interactions
between two huge markets, we may construct its internal structure
among {\it all} participants.

\section{Theoretical Background}
Let us consider two processes, $I$ and $J$. Transfer entropy \cite{Schrb2000}
from $J$ to $I$ is defined as follows:

\begin{equation}
\label{te1} T_{J \rightarrow I}=\sum p(i_{t+1}, i_{t}^{(k)},
j_{t}^{(l)}) \log \frac{p(i_{t+1}|i_{t}^{(k)},
j_{t}^{(l)})}{p(i_{t+1}|i_{t}^{(k)})}
\end{equation}

where $i_t$ and $j_t$ represent the states at time $t$ of $I$ and $J$,
respectively.. In terms of relative entropy, it can be rephrased as the
{\it distance} from the assumption that $J$ has no influence on $I$
(i.e. $p(i_{t+1}|i_{t}^{(k)}, j_{t}^{(l)})= p(i_{t+1}|i_{t}^{(k)})$).
One may rewrite Eq. (\ref{te1}) as:

\begin{equation}
\label{te2}
\begin{array}{rcl}
T_{J \rightarrow I} & = & H(i_{t+1}^{(k+1)}, j_{t}^{(l)}) - H(i_{t}^{(k)}, j_{t}^{(l)}) - H(i_{t+1}^{(k+1)}) + H(i_{t}^{(k)}) \\
& = & h_I (k;t) - h_{IJ} (k,l;t),
\end{array}
\end{equation}

from the property of conditional entropy. Then the second equality shows
that TE measures the change of entropy rate with knowledge of the process
$J$. Eq. (\ref{te2}) is practically useful, since the TE is decomposed into
entropy terms and there has been already well developed technique in entropy
estimation.

There are two choices in estimating entropy of a given time series.
First, the symbolic encoding method divides the range of the given dataset
into disjoint intervals and assign one symbol to each interval. The dataset,
originally continuous, becomes a discrete symbol sequence. Marschinski and
Kantz \cite{Mars2002} took this procedure and introduced the concept
called {\it effective transfer entropy}. The other choice exploits the
generalized correlation integral $C_q$. Prichard and Theiler
\cite{Prich1995} showed that the following holds for data $i$:

\begin{equation}
\label{corr} H_q (i, 2\epsilon) \approx -\log_2 [C_q (i, \epsilon)],
\end{equation}

where $\epsilon$ determines the size of a box in the box-counting algorithm.
We define the fraction of data points which lie within $\epsilon$ of $i(t)$ by

\begin{equation}
B(x(t),\epsilon) = \frac{1}{N} \sum_{s \neq t} \Theta (\epsilon-|i(t)-i(s)| ),
\end{equation}

where $\Theta$ is the Heaviside function, and calculate its numerical
value by the help of the box-assisted neighbor search algorithm
\cite{Kantz1997} after embedding the dataset into an appropriate phase space.
The generalized correlation integral of order $1$ is then given by

\begin{equation}
\log C_1 (i, \epsilon) = \frac{1}{N} \sum_{t} \log B(i(t), \epsilon).
\end{equation}

Notice that $H_q$ is expressed as an averaged quantity along the
trajectory $i(t)$ and it implies a kind of ergodicity which converts
an ensemble average $\sum_{i} p_k \log p_k = \langle \log p_k \rangle$
into a time average, $\overline{\log p_{k(t)}}$. Temporal correlations
are not taken into consideration since the daily data already lacks
much of its continuity.

It is rather straightforward to calculate entropy from a discrete
dataset using symbolic encoding. But determining the partition
remains as a serious problem, which is referred to as the {\it
generating partition problem}. Even for a two-dimensional
deterministic system, the partition lines may exhibit considerably
complicated geometry \cite{Grass1985,Christ1995} and thus should be
set up with all extreme caution \cite{Bollt2000}. Hence the
correlation integral method is often recommended if one wants to
handle continuous datasets without over-simplification, and we will
take this route. In addition, one has to determine the
parameter $\epsilon$. In a sense, this parameter plays a role of
defining the resolution or the scale of concerns, just as the number
of symbols does in the symbolic encoding method.

Before discussing how to set $\epsilon$, we remark on the finite
sampling effect: Though it is pointed out that the case of $q=2$
does not suffer much from finiteness of the number of data
\cite{Grass1983}, then the positivity of entropy is not guaranteed
instead \cite{Schrb2000}. Thus we choose the conventional Shannon
entropy, $q=1$ throughout this paper. There have been works done
\cite{Grass1988,Herzel1994,Roul1999} on correcting entropy
estimation. These correction methods, however, can be problematic
when calculating TE, since the fluctuations in each
term of Eq. (\ref{te2}) are not independent and should not be
treated separately \cite{Kaiser2002}. We actually found that a
proper selection of $\epsilon$ is quite crucial, and decided to
inactivate the correction terms here.

\begin{figure}
\includegraphics[angle=0,width=1.0\textwidth]{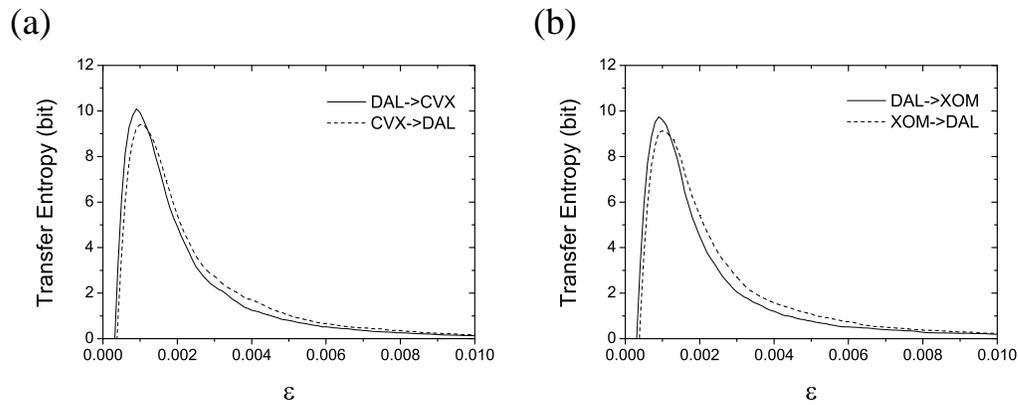}
\caption{Transfer entropy as function of $\epsilon$ between (a)
Delta Airlines (DAL) and Chevron (CVX) and (b) DAL and Exxon Mobil
(XOM).} \label{fig_te1}
\end{figure}

\begin{figure}
\includegraphics[angle=0,width=1.0\textwidth]{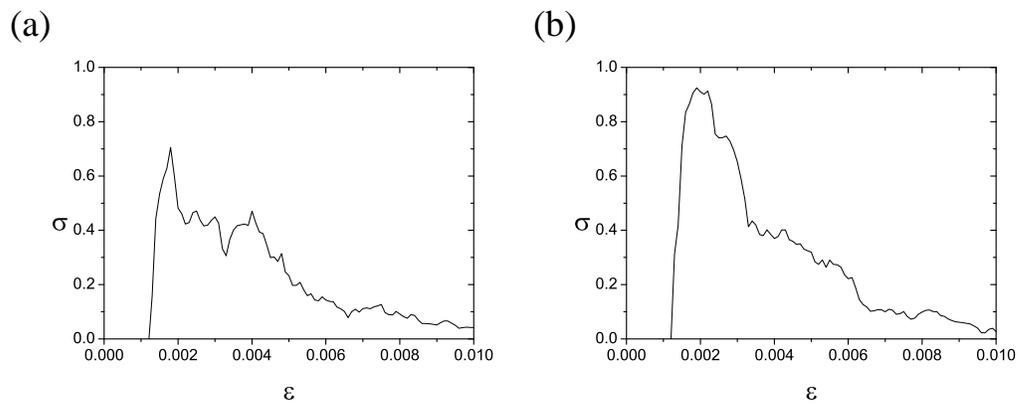}
\caption{Net information flow index, $\sigma$, between (a)DAL and
 CVX and (b)DAL and XOM} \label{fig_te2}
\end{figure}

A good value of $\epsilon$ will discriminate a real effect from
zero. Without {\it a priori} knowledge, we need to scan the range of
$\epsilon$ in order to find a proper resolution which yields
meaningful results from a time series. For reducing the
computational time, however, we resort to the empirical observation
that an airline company is quite dependent on the oil price while
the dependency hardly appears in the opposite direction.
Fig. \ref{fig_te1} shows this unilateral
effect: the major oil companies, Chevron and Exxon Mobile, have
influence over Delta Airline. From $\epsilon \simeq 0.002$ which
maximizes the difference between two directions (Fig.
\ref{fig_te2}), we choose the appropriate scale for analyzing the
data. Even in observing the temporal evolution, this value gives
good discrimination through the whole period. In Fig. \ref{fig_te1},
the influence seems reversed on very small length scales. The TE,
however, is known to increase monotonically under refinement of the
partitions in many cases \cite{Kaiser2002} and the refined partition
means the small length scale which is covered by the small $\epsilon$
in the correlation integral method. Hence we regard this
reversal as a finite sample effect in this paper, but it seems
worth looking further into the characteristics of TE analysis. And
we set $k=l=1$ in Eq. (\ref{te1}) since other values does not make
significant differences.

\section{Data Analysis}
This study deals with the daily closure prices of 135 stocks listed
on New York Stock Exchange (NYSE) from 1983 to 2003 ($L \simeq 5,000$
trading days, $\Delta t = 1$ trading day), obtained through the
website \cite{yahoo}. We select stocks which is listed on NYSE over
the whole periods. The companies in a stock market are usually
grouped into business sectors or industry categories, and our data
contain 9 business sectors (Basic Materials, Utilities, Healthcare,
Services, Consumer Goods, Financial, Industrial Goods,
Conglomerates, Technology) and 69 industry categories. The following
method shows how the information flows between the groups:

Suppose that we have a time series data $\{p(t)\}$, representing
the daily closure price of a company at time $t$. A stock market
analysis usually prefers treating the log return value:
\begin{equation}
i(t_2; t_1) = \log (\frac{p(t_2)}{p(t_1)})
\end{equation}
to the original price itself, since it satisfies the additive
property: $\sum_{k=0}^{N-1}i(t_{k+1}; t_k) = i(t_N; t_0)$. This log
return transformation also make the result invariant under the
arbitrary scaling of the input data. Therefore, in order to measure
the information transfer between two companies, say $I$ and $J$, we
create the log return time series $\{i(t)\}$ and $\{j(t)\}$ from the
raw price data. Then one can calculate the transfer entropies $T_{I
\rightarrow J}$ and $T_{J \rightarrow I}$ between them from the
equalities in the Section 2.

For obtaining an overview of the market, we consider groups of
similar companies. Let $I$ be a company of the group $\mathcal A$,
and $J$ be one of the group $\mathcal B$. The \emph{information flow
index} between these two groups is defined as a simple sum:

\begin{equation}
\rho_{\mathcal A \rightarrow \mathcal B}=\sum_{IJ}{T_{I \rightarrow J}} \label{information flow}.
\end{equation}

In addition, we define the \emph{net information flow index} to measure the
disparity in influences of the two groups as:

\begin{equation}
\sigma_{\mathcal A \mathcal B} = \rho_{\mathcal A \rightarrow
\mathcal B} - \rho_{\mathcal B \rightarrow \mathcal A}.
\end{equation}

If $\sigma_{\mathcal A \mathcal B}$ is positive, we can
say that the category $\mathcal A$ influences to the
category $\mathcal B$.

We examine the market with two grouping methods. One is business
sector, and the other is industry category. Grouping into business
sectors, however, does not exhibit clear directionality: the
influence of the $\mathcal A$ sector just alternates from that of
the $\mathcal B$ sector. In other words, the difference between
$\sigma_{\mathcal A \mathcal B}$ and $\sigma_{\mathcal B \mathcal
A}$ over the whole period is almost 0 (zero). This unclarity comes
from the fact that a business sector contains so many diverse
companies that its directionality just cancels out. On the other
hand, if we construct the asset tree through the minimum spanning
tree, each business sector forms a subset of the asset tree and the
subsets are connected mainly through the hub. Then, it can
be said that each of the business sectors forms a cluster
\cite{Onnela2003} and there are no significant direct links among them.

Hence we employ the industry category grouping, more detailed than the
business sectors. We have to exclude the categories which contain
only one element, and Table \ref{category} lists the remaining
industry categories used in the analysis.

\begin{figure}
\includegraphics[angle=0,width=1.0\textwidth]{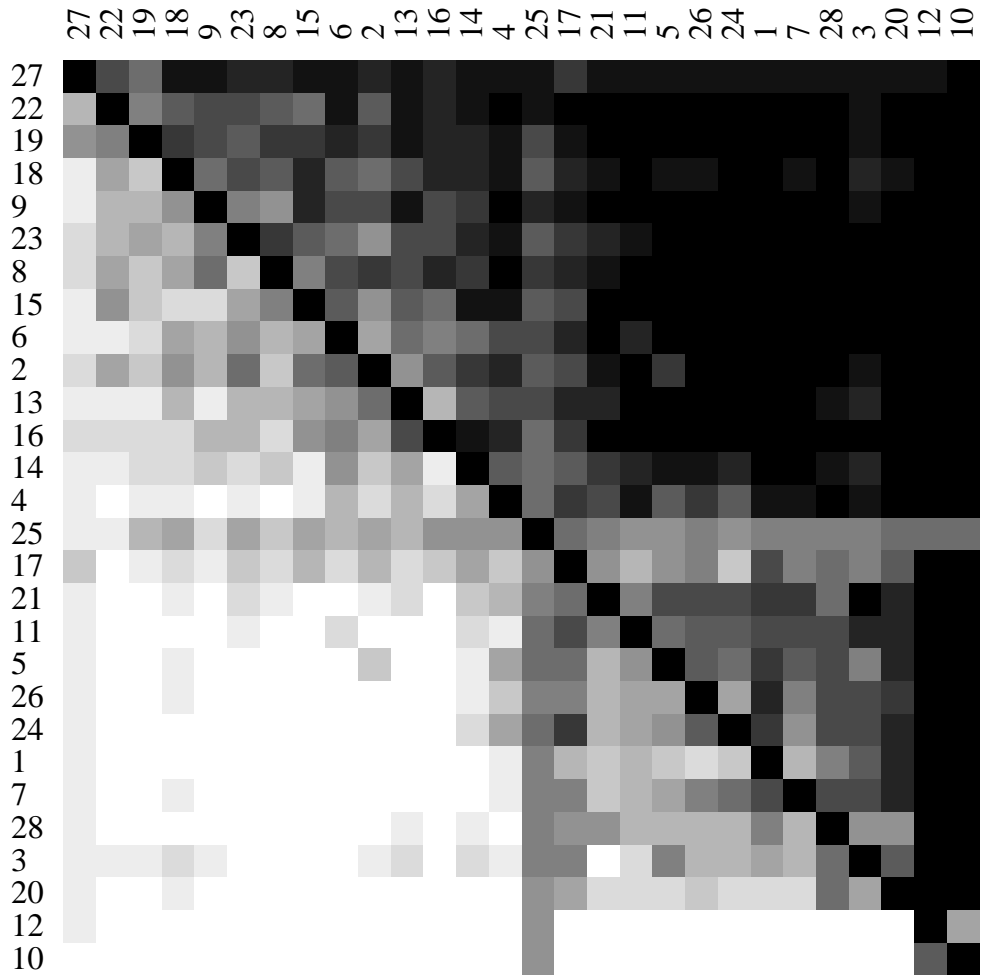}
\caption{$\sigma_{\mathcal A \mathcal B}$ over the whole 14 periods.
The degree of darkness represents the number of periods when $\mathcal A$
is affected by $\mathcal B$, and $\sigma_{\mathcal A \mathcal A}$'s are
left blank. For example, the category 10 affects almost all the other
categories and is affected by the categories 12 and 25 in a few periods.
The row of a market-leading category is bright on the average, while that
of a market-sensitive one is dark.} \label{map}
\end{figure}

As in our previous observation, it is verified again that oil
companies and airline companies are related in a unilateral way: The
category 20, Major Oil \& Gas, has continuing influence over the
category 19, Major Airline, during the whole 14 periods under
examination ($\sigma_{20,19} > 0$) . One can easily find
such relations in other categories: for example, the category 20
always influences on the categories 15 (Independent Oil\&Gas), 22
(Oil\&Gas Equipment\&Services), and 23 (Oil\&Gas
Refining\&Marketing). It also affects the category 27 (Regional
Airlines) over 13 periods and maintains its power on the whole
market during 11 periods (Fig. \ref{map}).

It is well-known that economy greatly depends on the energy supply
and price such as oil and gas. Transfer entropy analysis
quantitatively proves this empirical fact. The top three influential
categories (in terms of periods) are the categories 10 (Diversified
Utilities), 12 (Electric Utilities) and 20. All of ten companies in
the categories 10 and 12 are again related to the energy industry,
such as those for holding, energy delivery, generation,
transmission, distribution, and supply of electricity.

On the contrary, an airline company is sensitive to the tone of the
market. These companies receive information from other categories
almost all the time (category 19: 11 periods, category 27: 12
periods). The category 8 (Credit Services) and the category 9
(Diversified Computer Systems, including only HP and IBM in our
data) are also market-sensitive as easily expected.

\section{Conclusion}
We calculated the transfer entropy with the daily data of the US
market. The concept of transfer entropy provides a quantitative
value of general correlation and the direction of information. Thus
it reveals how the information flows among companies or groups of
companies, and discriminates the market-leading companies from the
market-sensitive ones. As commonly known, the energy such as natural
resources and electricity is shown to greatly affect economic
activities and the business barometer. This analysis may be applied to
predicting the stock price of a company influenced by other ones. 

In short, TE proves its possibility as a promising measure to detect
directional information. We suggest that the merits and demerits of
TE should be judged in details with respect to those of the
classical methods like the correlation matrix theory.

\begin{ack}
We thank Dong-Hee Kim, Hawoong Jeong and Jae-Suk Yang for helpful
discussion.
\end{ack}


\newpage
\clearpage

\begin{table}
\caption{Industry category index in alphabetic order}
\label{category}
\begin{tabular}{|c|c|}
\hline
\#&industry category\\
\hline
1&Aerospace/Defense\\
2&Auto Manufacturers\\
3&Beverages\\
4&Business Equipment\\
5&Chemicals\\
6&Communication Equipment\\
7&Conglomerates\\
8&Credit Services\\
9&Diversified Computer Systems\\
10&Diversified Utilities\\
11&Drug Manufacturers\\
12&Electric Utilities\\
13&Farm\&Construction Machinery\\
14&Health Care Plans\\
15&Independent Oil\&Gas\\
16&Industrial Metals\&Minerals\\
17&Information Technology Services\\
18&Lumber, Wood Production\\
19&Major Airlines\\
20&Major Oil\&Gas\\
21&Medical Instruments\&Supplies\\
22&Oil\&Gas Equipment\&Services\\
23&Oil\&Gas Refining\&Marketing\\
24&Personal Products\\
25&Processing Systems\&Products\\
26&Railroads\\
27&Regional Airlines\\
28&Specialty Chemicals\\
29&Etc.\\
\hline
\end{tabular}
\end{table}

\end{document}